\documentstyle[12pt,a4,epsf]{article}
\begin{document}
\title{Two-dimensional Burgers Cellular Automaton} 
\author{
Katsuhiro Nishinari, Junta Matsukidaira\\
Department of Applied Mathematics and Informatics, \\
Ryukoku University, Seta, Ohtsu 520-2194, JAPAN\\
and \\
Daisuke Takahashi\\
Department of Mathematical Sciences, \\
Waseda University, Ohkubo, Shinjuku-ku, Tokyo 169-8555, JAPAN \\}
\date{}
\maketitle
\begin{abstract}
In this paper, a two-dimensional cellular automaton(CA) associated with
a two-dimensional Burgers equation is presented.
The 2D Burgers equation is an integrable generalization of the well-known
Burgers equation, and is transformed into a 2D
diffusion equation by the Cole-Hopf transformation.
The CA is derived from the 2D Burgers equation by using
the ultradiscrete method, which can transform dependent variables into
 discrete ones.
Some exact solutions of the CA, such as shock wave solutions, are
 studied in detail. 
\end{abstract}
\newpage
\section{Introduction}
Recently, cellular automaton(CA) is extensively studied
as a method to make models of complex systems, such as 
traffic flow\cite{NS}, fluid dynamics\cite{MZ}, immune
systems\cite{CS}.
Since CA models are represented by binary procedures
or discrete equations, 
exact analysis or explicit solutions are rarely obtained.
Thus we often use numerical calculations or
thermodynamical and statistical methods\cite{SSNI}\cite{DEHP}
to study global features.

Recently, the ultradiscrete method is proposed as a new type of
discretization technique which allows us to study direct 
relations between continuous system and CA\cite{TTMS}\cite{Matsu}.
As for soliton equations, we can derive associated CA from
a continuous equation keeping their
integrable properties by the method.
Another example of applying this method 
has been given by two of the authors and 
a CA associated with the 1D Burgers equation is derived as
follows\cite{NT1}:
\begin{equation}
 U^{t+1}_i=U^t_i+\min(U^t_{i-1},L-U^t_i)-\min(U^t_i,L-U^t_{i+1}).
\label{oriBCA}
\end{equation}
The CA is called Burgers CA(BCA), and it has nice properties;
it can be transformed into a linear CA of
diffusion type, and 
is a multi-value generalization of the rule--184 CA\cite{Wo}. 
Moreover, BCA and its extensions can be used as highway traffic models, 
and properties of models and observed data are 
compared in detail\cite{NT2}\cite{NT3}.

In this paper, we generalize the previous results to two spatial
dimension.
First, we introduce 2D Burgers equation, which 
can be linearized by a dependent variable transformation.
Then we discretize independent variables keeping its integrable
properties. Finally, we use the ultradiscrete method 
to obtain two-dimensional BCA. 
We also study some exact solutions of the CA. 
\section{Two-dimensional Burgers equation and its ultradiscretization}
2D Burgers equation we consider in this paper is
\begin{eqnarray}
 u_t&=&u_{xx}+u_{yy}+2uu_x+2vu_y,\label{cBS1}\\
 v_x&=&u_y,\label{cBS2}
\end{eqnarray}
or, eliminating the variable $v$, we can rewrite these into a single equation
\begin{equation}
 u_t=u_{xx}+u_{yy}+2uu_x+2u_y\int u_y dx.
\label{cBSu}
\end{equation}
This equation has a remarkable property that it can be linearized
by the Cole--Hopf transformation
\begin{equation}
 u=\frac{f_x}{f}
\label{cCH}
\end{equation}
into a 2D diffusion equation
\begin{equation}
 f_t=f_{xx}+f_{yy}.
\label{2Ddiff}
\end{equation}
Thus the Burgers equations are integrable in a sense that they can be
transformed into a linear equation.
It is noted that from (\ref{cBS2}) and (\ref{cCH}), we obtain 
\begin{equation}
  v=\frac{f_y}{f}.
\label{cCHv}
\end{equation}
From (\ref{cCH}) and (\ref{cCHv}) it is apparent that the variables
$u$ and $v$ can be treated symmetrically, 
then we can derive the time
evolution equation for $v$.
From (\ref{2Ddiff}) and (\ref{cCHv}), we obtain 
\begin{equation}
 v_t=v_{xx}+v_{yy}+2vv_y+2uv_x.
\label{cBSv}
\end{equation} 
Thus we can write 2D Burgers equation in another
coupled form as
\begin{equation}
 {\bf u}_t=\nabla^2 {\bf u}+2({\bf u}\cdot \nabla){\bf u},
\end{equation}
where ${\bf u}=(u,v)$.
This form appears in \cite{Ku}, in which it is called
vector Burgers equation. 

To discretize the independent variables in (\ref{cBS1}) and (\ref{cBS2}), 
we utilize discrete analogues to (\ref{cCH}) and (\ref{2Ddiff})\cite{NT1}.
Discretizing both time and space variables in (\ref{2Ddiff}), 
we obtain a discrete diffusion equation
\begin{equation}  \label{discdiff}
  f_{ij}^{t+1}=(1-4\delta)f_{ij}^{t}+\delta(f_{i+1j}^{t}+f_{i-1j}^{t}+
                                           f_{ij+1}^{t}+f_{ij-1}^{t})
\end{equation}
where $\delta=\Delta t/h^2$ and $h=\Delta x=\Delta y$.
$\Delta t, \Delta x$ and $\Delta y$ are lattice intervals 
in time $t$, space $x$ and $y$, respectively.  
Next we define discrete analogues to the Cole--Hopf transformation
(\ref{cCH}) and (\ref{cCHv})
\begin{equation}  \label{dCH}
  u_{ij}^t \equiv c\frac{f_{i+1j}^t}{f_{ij}^t},\,\,\,  
  v_{ij}^t \equiv c\frac{f_{ij+1}^t}{f_{ij}^t}
\end{equation}
where $c$ is a constant.  
Evolution equations for $u^t_{ij}$ and $v^t_{ij}$ are derived from 
(\ref{discdiff}) and (\ref{dCH}) as
\begin{eqnarray}
  u_{ij}^{t+1} &=& u_{ij}^{t}
    \frac{\displaystyle{
     \frac{1-4\delta}{c\delta} + \frac{u_{i+1j}^t}{c^2}+ \frac{1}{u_{ij}^t}
                               + \frac{v_{i+1j}^t}{c^2}+ \frac{1}{v_{i+1j-1}^t}
    }}{\displaystyle{
     \frac{1-4\delta}{c\delta} + \frac{u_{ij}^t}{c^2}+ \frac{1}{u_{i-1j}^t}
                               + \frac{v_{ij}^t}{c^2}+ \frac{1}{v_{ij-1}^t}
    }},\label{dBS1}\\
\frac{v^t_{i+1j}}{v^t_{ij}}&=&\frac{u^t_{ij+1}}{u^t_{ij}}.\label{dBS2}
\end{eqnarray}
If we take an appropriate continuous limit $\Delta t, h\to 0$,
we obtain (\ref{cBS1}) and (\ref{cBS2}) from (\ref{dBS1}) and
(\ref{dBS2}).
Moreover, (\ref{dBS1}) and (\ref{dBS2}) can be transformed into the
linear diffusion equation (\ref{discdiff}) through (\ref{dCH}).

Next, we ultradiscretize (\ref{dBS1}) and (\ref{dBS2}), 
that is, discretize dependent variables $u$ and $v$.
Let us introduce a transformation of variables 
and parameters as follows:
\begin{eqnarray}
   u_{ij}^t &=& e^{U_{ij}^t/\varepsilon},\,\,\,\,
   v_{ij}^t = e^{V_{ij}^t/\varepsilon},  \label{uU}\\
   \frac{1-4\delta}{c\delta} &=& e^{-M/\varepsilon},  \label{dM}\\
   c^2 &=& e^{L/\varepsilon},  \label{cL}
\end{eqnarray}
where $\varepsilon$ is a parameter.
Substituting these transformations into (\ref{dBS1}) and (\ref{dBS2}),
taking a limit  $\varepsilon\to+0$
and using the formula
\begin{equation}
 \lim_{\varepsilon\to+0}\varepsilon\log(\exp(\frac{A}{\varepsilon})
+\exp(\frac{B}{\varepsilon}))=\max(A,B)=-\min(-A,-B),
\end{equation}
we obtain 
\begin{eqnarray}  
  U_{ij}^{t+1} &=& U_{ij}^t + \min(M, U_{i-1j}^t, L - U_{ij}^t, 
                                                V_{ij-1}^t, L - V_{ij}^t)
\nonumber\\
                   && - \min(M, U_{ij}^t, L - U_{i+1j}^t, 
                                                V_{i+1j-1}^t, L - V_{i+1j}^t),
\label{uBS1}\\
 V_{i+1j}^t-V_{ij}^t&=& U_{ij+1}^t-U_{ij}^t.\label{uBS2}
\end{eqnarray}
These are two-dimensional generalization of BCA (\ref{oriBCA}).
We call this system two-dimensional Burgers CA(2DBCA).
In (\ref{uBS1}), we can eliminate the variable $V$ by using 
(\ref{uBS2}) and obtain
\begin{equation}
   U_{ij}^{t+1} = U_{ij}^t + Q_{ij}^t - Q_{i+1j}^t,
\label{uBSu}
\end{equation}
where flow $Q_{ij}^t$ is given by
\begin{equation}
 Q_{ij}^t = \min(M, U_{i-1j}^t, L - U_{ij}^t,V_{0j-1}^t+
\sum_{k=0}^{i-1}(U^t_{kj}-U^t_{kj-1}),L-V_{0j}^t-
\sum_{k=0}^{i-1}(U^t_{kj+1}-U^t_{kj})),
\label{flowqq}
\end{equation}
which is an ultradiscrete version of (\ref{cBSu}).
In (\ref{flowqq}), $V_{0j}^t$ represents boundary conditions for $V$
at $i=0$.
Since variables $u$ and $v$ can be treated symmetrically, we can
derive an equation for $V$ by using 
a similar procedure to the above.
The result is
\begin{eqnarray}  
  V_{ij}^{t+1} = V_{ij}^t &+& \min(M, U_{i-1j}^t,\,L-U_{ij}^t,\,
                                     V_{ij-1}^t,\,L-V_{ij}^t) 
\nonumber\\
    &-& \min(M, U_{i-1j+1}^t,\,L-U_{ij+1}^t,\,V^t_{ij},\,L-V_{ij+1}^t).
\label{anoV}
\end{eqnarray}
It should be noted that 
2DBCA is also expressed by a set of equations
(\ref{uBS1}) and (\ref{anoV}),
and it is equivalent to that of (\ref{uBS1}) and (\ref{uBS2})
under an appropriate condition. 
This fact is directly shown as follows.
Let us define $h^t_{ij}$ by
\begin{equation}
  h^t_{ij} \equiv \min(M, U_{ij}^t,\,L-U_{i+1j}^t,\,V^t_{i+1j-1},\,L-V_{i+1j}^t),
\end{equation}
then (\ref{uBS1}) and (\ref{anoV}) are written as 
\begin{eqnarray}
  U_{ij}^{t+1} &=& U_{ij}^t + h^t_{i-1j} - h^t_{ij} \\
  V_{ij}^{t+1} &=& V_{ij}^t + h^t_{i-1j} - h^t_{i-1j+1}.
\end{eqnarray}
By using these equation we have 
\begin{eqnarray}
&&V_{i+1j}^{t+1} - V_{ij}^{t+1} - U_{ij+1}^{t+1} + U_{ij}^{t+1} \nonumber\\
&=& \{V_{i+1j}^t + h^t_{ij} - h^t_{ij+1}\}
  - \{V_{ij}^t + h^t_{i-1j} - h^t_{i-1j+1}\} \nonumber\\
&& \qquad - \{U_{ij+1}^t + h^t_{i-1j+1} - h^t_{ij+1}\}
  + \{U_{ij}^t + h^t_{i-1j} - h^t_{ij}\} \nonumber\\
&=& V_{i+1j}^t - V_{ij}^t - U_{ij+1}^t + U_{ij}^t.
\end{eqnarray}
Thus 
\begin{equation}
  V_{i+1j}^t - V_{ij}^t - U_{ij+1}^t + U_{ij}^t = \hbox{const.}
\end{equation}
holds for any time $t$.
By choosing initial condition of $U$ and $V$ so that the above constant
becomes zero everywhere, then (\ref{uBS2}) is always satisfied.
Note that if $U$ is periodic in lattice $i$ and $V$ in $j$, then
the total sums $\sum_{i} U^t_{ij}$ and $\sum_{j} V^t_{ij}$ are
conserved quantities, which is directly shown by using (\ref{uBS1}) and
(\ref{anoV}).

Finally, we consider a relation between 
the ultradiscrete Burgers equation and ultradiscrete diffusion equation.
Introducing a transformation
\begin{equation} 
f_{ij}^t = \exp(F_{ij}^t/\varepsilon),
\label{fF}
\end{equation}
ultradiscrete Cole--Hopf transformations
\begin{eqnarray} 
  U_{ij}^t &=& F_{i+1j}^t - F_{ij}^t + {L\over2},\label{uUch}\\
  V_{ij}^t &=& F_{ij+1}^t - F_{ij}^t + {L\over2}\label{uVch}
\end{eqnarray}
are obtained from (\ref{dCH}) and (\ref{uU}) under the limit $\varepsilon\to+0$.  
Substituting (\ref{fF}) into (\ref{discdiff}),
 we obtain an ultra-discrete diffusion equation;
\begin{equation} \label{uldiff1}
  F_{ij}^{t+1} = \max(F_{ij}^t+{L\over2}-M, F_{i+1j}^t, F_{i-1j}^t, 
                              F_{ij+1}^t, F_{ij-1}^t)-
                 \max(0,{L\over2}-M).
\end{equation}
This is also obtained as follows. If we substitute (\ref{uUch}) and
(\ref{uVch}) into (\ref{uBS1}), we have
\begin{eqnarray}
 F_{i+1j}^{t+1} - F_{ij}^{t+1}&=&
\max(F_{ij}^t+{L\over2}-M, F_{i+1j}^t, F_{i-1j}^t, 
                              F_{ij+1}^t, F_{ij-1}^t)\nonumber\\
&-&\max(F_{i+1j}^t+{L\over2}-M, F_{i+2j}^t, F_{ij}^t,
                              F_{i+1j+1}^t, F_{i+1j-1}^t).
\end{eqnarray}
Since this equation is in a recurrence form in $i$, we obtain
(\ref{uldiff1})
if we choose the decoupling constant in the above equation
as $-\max(0,{L/2}-M)$.
This shows that 
we can linearize 2DBCA by the ultradiscrete Cole-Hopf transformations and
the CA can be considered to keep integrable property.
(\ref{uBS2}) is automatically satisfied by substituting 
(\ref{uUch}) and (\ref{uVch}) into it, and this means that (\ref{uBS2})
represents the compatibility condition of differences in $i$ and $j$ 
direction of the variable $F$.
Note that we consider a linearity of (\ref{uldiff1}) in a sense that
superposition of solutions is realized by $\max$ 
operation like '+' in (\ref{2Ddiff}).
\section{Some solutions of 2DBCA}
\subsection{Shock wave solution}
In this section, we discuss some solutions and examine their behavior in
detail. First, we derive an exact solution which represents shock waves.
In order to obtain the solution, we utilize the shock wave solution
of the discrete Burgers equation (\ref{dBS1}) and (\ref{dBS2}).
Let us assume $f_{ij}^t$ has the following form
\begin{equation}  \label{Nshock}
  f_{ij}^t = 1 + \sum_{z=1}^{N}\exp(k_zi + r_zj + \omega_z t + \xi_z),
\end{equation}
where $k_z$, $r_z$, $\omega_z$ and $\xi_z$ $(z=1,2,\cdots,N)$ are constants.  
(\ref{Nshock}) is an exact solution of (\ref{discdiff})
under the condition
\begin{equation}  \label{disper}
  \omega_z = \log(1 -4\delta + \delta(e^{k_z} + e^{-k_z} + e^{r_z}+ e^{-r_z}))
   \,\,\,\,\, (z=1,2,\cdots,N),
\end{equation}
which is obtained simply substituting (\ref{Nshock}) into (\ref{discdiff}). 
(\ref{disper}) is the dispersion relation between the 
frequency ($\omega_z$) and the wavenumber ($k_z$ and $r_z$).
From (\ref{dCH}) we have
\begin{equation}
 u_{ij}^t=c\frac{1 + \sum_{z=1}^{N}\exp(k_z(i+1) + r_zj + \omega_z t + \xi_z)}{
1 + \sum_{z=1}^{N}\exp(k_zi + r_zj + \omega_z t + \xi_z)},
\label{NSU}
\end{equation}
and also have $v^t_{ij}$.
We call this
``$N$-shock wave'' solution of (\ref{dBS1}) and (\ref{dBS2}).
From this solution we can obtain a shock wave solution of 
2DBCA by the ultradiscrete method.
First we assume
\begin{equation}
  k_z = \frac{K_z}{\varepsilon},\ r_z = \frac{R_z}{\varepsilon},\ 
  \omega_z = \frac{\Omega_z}{\varepsilon},\ 
  \xi_z = \frac{\Xi_z}{\varepsilon},
\end{equation}
and noticing (\ref{uU}) and (\ref{cL}), we obtain from (\ref{NSU})
\begin{eqnarray} 
  U_{ij}^t &=& \frac{L}{2} + \max(0, \theta_1(i+1,j,t), \theta_2(i+1,j,t), 
\cdots , \theta_N(i+1,j,t))\nonumber\\
           &-& \max(0, \theta_1(i,j,t), \theta_2(i,j,t), \cdots , \theta_N(i,j,t)),
\label{nswsol}
\end{eqnarray}
by taking the limit $\varepsilon\to+0$. Here phases $\theta_z
(z=1,\cdots,N),$ are given by
\begin{equation}
 \theta_z(i,j,t)=K_z i + R_zj + \Omega_z t + \Xi_z.
\label{swphase}
\end{equation}
From (\ref{disper}), a dispersion relation becomes in the limit
\begin{equation}
  \Omega_z = \max(|K_z|,|R_z|,\frac{L}{2}-M)-\max(0,\frac{L}{2}-M).
\end{equation}
(\ref{nswsol}) is an exact solution of (\ref{uBS1}) and 
we also get a solution $V^t_{ij}$ by (\ref{uBS2}),
which represent an ultradiscrete shock wave.
The $\max$ operation in (\ref{nswsol}) represents the superposition
principle, as mentioned in the previous section.

Fig.1 shows an example of shock wave solution (\ref{nswsol}) in the case of $N=2$.
Parameters are taken as $K_1=-1, R_1=K_2=R_2=1$, $\Xi_1=\Xi_2=0$,
$L=2$ and $M=2$.
We easily see that 
$\displaystyle{\lim_{i\to\infty}}U^t_{ij}= L/2+K_2=2$, and
$\displaystyle{\lim_{i\to-\infty}}U^t_{ij}=L/2+K_1=0$,
and $\displaystyle{\lim_{j\to-\infty}}U^t_{ij}=L/2=1$.
\subsection{Particle model and excitation behavior}
If we assume the parameter $M$ is sufficiently larger than $L/2$, 
we can neglect terms that contain $M$ in (\ref{uBS1}), (\ref{anoV})
and (\ref{uldiff1}), and simplify the analysis on them.
In this case equations on $U$ and $V$ become 
\begin{eqnarray}  
  U_{ij}^{t+1} &=& U_{ij}^t + \min(U_{i-1j}^t, L - U_{ij}^t, 
                                                V_{ij-1}^t, L - V_{ij}^t)
\nonumber\\
                  &&- \min(U_{ij}^t, L - U_{i+1j}^t, 
                                                V_{i+1j-1}^t, L - V_{i+1j}^t),
\label{UMnot}\\
  V_{ij}^{t+1} &=& V_{ij}^t + \min(U_{i-1j}^t,\,L-U_{ij}^t,\,
                                     V_{ij-1},\,L-V_{ij}^t) 
\nonumber\\
    &&- \min(U_{i-1j+1}^t,\,L-U_{ij+1}^t,\,V_{ij},\,L-V_{ij+1}^t),
\label{VMnot}
\end{eqnarray}
and equation on $F$ becomes
\begin{equation}  \label{uldiff2}
  F_{ij}^{t+1} = \max(F_{i+1j}^t, F_{i-1j}^t,F_{ij+1}^t, F_{ij-1}^t).
\end{equation}
It is easily verified that if initial conditions are taken
as $0\le F_{ij}^0, 0\le U_{ij}^0\le L$ and $0\le V_{ij}^0\le L$, then
these conditions hold for any time $t$.

The set of equations (\ref{UMnot}) and (\ref{VMnot}) 
express a simple particle model.
Before explaining the model, we shortly review a one-dimensional
particle model defined by (\ref{oriBCA}) described in \cite{NT1}.
In the model, there are one-dimensional infinite sites indexed by $i$.
Every site can hold $L$ particles at most.
Particles at each site move to empty spaces in its neighboring right
site synchronously at each time step.
According to this rule, flow from site $i$ to $i+1$ at time $t$
becomes $\min(U^t_i,L-U^t_{i+1})$ if $U^t_i$ denotes the number of
particles at site $i$ and time $t$.
Thus considering the number of particles coming into and 
escaping from site $i$, we obtain (\ref{oriBCA}).

Next we explain two-dimensional particle model expressed by 
(\ref{UMnot}) and (\ref{VMnot}).
The model is constructed as follows:
\begin{enumerate}
 \item[(a)]There are two-dimensional infinite sites indexed by $i$ and
	   $j$.
 \item[(b)]There are two kinds of particles, shortly saying $U$-particles
	   and $V$-particles. $U$-particles move only in positive $i$ 
           direction. $V$-particles in positive $j$ direction, as shown 
           in Fig.2.
 \item[(c)]Every site can hold $L$ particles at most of both kinds
	   respectively.
 \item[(d)]$U$-particles at site $(i,j)$ can move to empty spaces in its 
           neighboring right site $(i+1,j)$ per unit time. $V$-particles 
           at $(i,j)$ can move to those in up site $(i,j+1)$.
 \item[(e)]The number of $U$-particles from $(i-1,j)$ to $(i,j)$ from
	   $t$ to $t+1$ must be equal to that of $V$-particles from 
           $(i,j-1)$ to $(i,j)$ at the same time step.
 \item[(f)]Under this restriction, $U$- and $V$-particles move to occupy 
           neighboring empty spaces.
\end{enumerate}
Let us define the number of $U$- and $V$-particles at site $(i,j)$ and time
$t$ by $U^t_{ij}$ and $V^t_{ij}$ respectively.
Then, according to the above rule (e) and (f),
flow of $U$-particles from $(i-1,j)$ to $(i,j)$ and that of $V$-particles
from $(i,j-1)$ to $(i,j)$ are both $\min(U^t_{i-1j},L-U^t_{ij},V^t_{ij-1},L-V^t_{ij})$.
Thus, evolution equations
of the above model become (\ref{UMnot}) and (\ref{VMnot}).
We call this rule ``pairing
rule'' because the same number of $U$- and $V$-particles come into the
same site.

We show simple examples of time evolution of the above model. 
We choose $L=2$ for all examples below.
As the first example, we set $F^0_{ij}=0$
except $F^0_{0,0}=1$ at the origin.
This means $U^0_{-1,0}=2$, $U^0_{0,0}=0$, $V^0_{0,-1}=2$,
$V^0_{0,0}=0$ and $U^0_{ij}=V^0_{ij}=1$ at all other points.
The time evolution is given in Fig.3.
An exact solution of (\ref{uldiff2}) is
\begin{equation}  \label{uldiffsmp}
  F_{ij}^t = \max_{{|k+l|\le t\atop|k-l|\le t}\atop k+l=t {\rm\ mod\ } 2}
                 G_{i+i_0-k,\,j+j_0-l},
\end{equation}
where $G_{k,l}$ is defined by
\begin{eqnarray}
  G_{k,l}&=& \min(\max(0,2k+1)-2\max(0,2k)+\max(0,2k-1), \nonumber\\
         && \qquad \max(0,2l+1)-2\max(0,2l)+\max(0,2l-1)),  \label{ultraG}
\end{eqnarray}
and $i_0$, $j_0$ is real phase constants.  Since $i_0$ and $j_0$ are arbitrary
real constants, the above solution satisfies (\ref{uldiff2}) even if $i$--$j$
lattice is shifted as $i\to i+i_0'$ and $j\to j+j_0'$.  If we set $i_0=j_0=0$
in (\ref{uldiffsmp}), we obtain the same solution shown in Fig.~3 on integer
lattice points.

We can derive the above solution from a solution of difference equation
(\ref{discdiff}) using ultradiscretization.
The derivation is shown in Appendix.
The solution represents an expansion of excited wave of $F$ as shown in Fig.3(a).
The solutions for $U$ and $V$ are obtained by substituting (\ref{uldiffsmp})
into (\ref{uUch}) and (\ref{uVch}), respectively, or 
simply by moving $U$-particles and $V$-particles according to the above
rule.
Fig.4 is an example of expanding rings of $U$ and $V$. 
Initial conditions are $F^0_{i,j}=0$ except $F^0_{0,0}=1$ and $F^0_{1,0}=1$.
The solution is apparently obtained by a superposition of (\ref{uldiffsmp}).
We see that $\sum_{i} U^t_{ij}$ and 
$\sum_{j} V^t_{ij}$ are constant for time in both cases.
\section{Concluding discussions}
In this paper, two-dimensional CA associated with
Burgers equation(2DBCA) is presented by using ultradiscrete method.
2DBCA has some remarkable properties that it can be
linearized by means of the ultradiscrete Cole--Hopf transformation,
and has exact solutions which shows $N$-shock wave and excitation
behaviors. 

We also show a particle model of the CA, which can be
considered as a pairing movement of particles in $i$ and $j$ direction.
We hope that this rule and its integrability contribute
to make models on some multi-dimensional complex systems.\\\\
\centerline{\bf {Acknowledgment}}
The authors are grateful to Prof. Ryogo Hirota and Prof. Yasuhiro Ohta
for fruitful discussions and helpful comments. 
This work is partially supported by Grant-in-Aid from the Ministry 
of Education, Science and Culture.
\newpage

\newpage
\begin{flushleft}
{\large {\bf Appendix}} 
\end{flushleft}
In this appendix, we show that (\ref{uldiffsmp}) can be derived from
a solution of the difference equation (\ref{discdiff}) through the
ultradiscretization.\par
  First, since we consider the case that the parameter $M$ is sufficiently
large, we can set $\delta=1/4$ by using (\ref{dM}).
Then (\ref{discdiff}) becomes
\begin{equation}  \label{discdiff4}
  f_{ij}^{t+1}=\frac14(f_{i+1j}^{t}+f_{i-1j}^{t}+
                                           f_{ij+1}^{t}+f_{ij-1}^{t}).
\end{equation}
We can interpret this equation as 2D random walk process.
Therefore, we obtain an exact solution including 2D binomial distribution,
\begin{equation}  \label{binomial}
  f_{ij}^t = \sum_{{|k+l|\le t\atop|k-l|\le t}\atop k+l=t {\rm\ mod\ } 2}
             c_{k,l}^t\,g_{i+i_0-k,\,j+j0-l},
\end{equation}
where
\begin{eqnarray*}
  && c_{k,l}^t = \frac{1}{4^t}\cdot\frac{t!}{((t-|k+l|)/2)!((t+|k+l|)/2)!}
             \cdot\frac{t!}{((t-|k-l|)/2)!((t+|k-l|)/2)!} \\
  && g_{k,l} = 1\Big/\Bigl\{\frac{(1+\exp(2k/\varepsilon))^2}
             {(1+\exp((2k+1)/\varepsilon))(1+\exp((2k-1)/\varepsilon))} \\
  && \qquad\qquad\qquad
    + \frac{(1+\exp(2l/\varepsilon))^2}
           {(1+\exp((2l+1)/\varepsilon))(1+\exp((2l-1)/\varepsilon))}\Bigr\}
\end{eqnarray*}
and $i_0$, $j_0$ is real phase constants, $\varepsilon$ is a limiting
parameter.  We have
$$
  \lim_{\varepsilon\to+0} \varepsilon\log g_{k,l}=G_{k,l}
        \ {\rm in}\ (\ref{ultraG})
$$
and $c_{k,l}^t$ is always positive finite for finite $k$, $l$ and $t$.
Therefore, taking a limit
$\displaystyle\lim_{\varepsilon\to+0}\varepsilon\log f_{ij}^t$,
all $c_{k,l}^t$ terms disappear and only $g_{k,l}$ survive in the summation
of (\ref{binomial}).  Thus we obtain $F_{ij}^t$ in (\ref{uldiffsmp}) from
$f_{ij}^t$ in (\ref{binomial}) by the ultradiscretization.  Figure A1 (a) shows
$F_{ij}^t$ , and (b) and (c) show $\displaystyle\varepsilon\log f_{ij}^t$.
We draw all figures as continuous plot on $i$ and $j$ and set $t=2$ and
$i_0=j_0=0$.  Parameter $\varepsilon$ is $0.2$ and $0.01$ in (b) and (c),
respectively.  From these figures, we can see
a difference solution converges to a ultradiscrete one as $\varepsilon\to+0$.
\newpage
\centerline{\bf {Figure Captions}}
\begin{enumerate}
\item[Fig.1]  The snapshot of shock wave solution at $t=1$ in the case of $N=2$.
Parameters are $K_1=-1, R_1=K_2=R_2=1$, $\Xi_1=\Xi_2=0$,
$L=2$ and $M=2$.
\item[Fig.2]  A physical explanation of the (2+1)-dimensional BCA,
	      called ``pairing rule''. The same number of $U$-particles 
and $V$-particles come into the same site. The solid arrow and broken
	      arrow show the movement of $U$ and $V$, respectively.
\item[Fig.3]  Excitation behavior of (a)$F$, (b)$U$, (c)$V$ at $t=6$. We choose $L=2$,
and initial conditions are $F^0_{i,j}=0$ except the origin $F^0_{0,0}=1$.
The light gray, dark gray and black square represent 0, 1, 2, respectively.
\item[Fig.4]  Expanding wave of $U$ and $V$ ((a)$F$, (b)$U$, (c)$V$ at
	      $t=6$). Initial conditions are $F^0_{i,j}=0$ except
	      $F^0_{0,0}=1$ and $F^0_{1,0}=1$. 
The light gray, dark gray and black square represent 0, 1, 2, respectively.
\item[Fig.A1]  Plot of solutions of ultradiscrete and difference diffusion
equation.
(a) $F_{ij}^t$ in (\ref{uldiffsmp}), (b) $\varepsilon\log f_{ij}^t$ in
(\ref{binomial}) for $\varepsilon=0.2$ and (c) $\varepsilon\log f_{ij}^t$ in
(\ref{binomial}) for $\varepsilon=0.01$.  In all figures, we set $t=2$ and
$i_0=j_0=0$.
\end{enumerate}
\newpage
\Large{
 \begin{center}
   {\bf Figure 1 by K.~Nishinari, et al}\\
 \end{center}
   \vspace{6cm}
   \epsfxsize=15cm \epsfbox{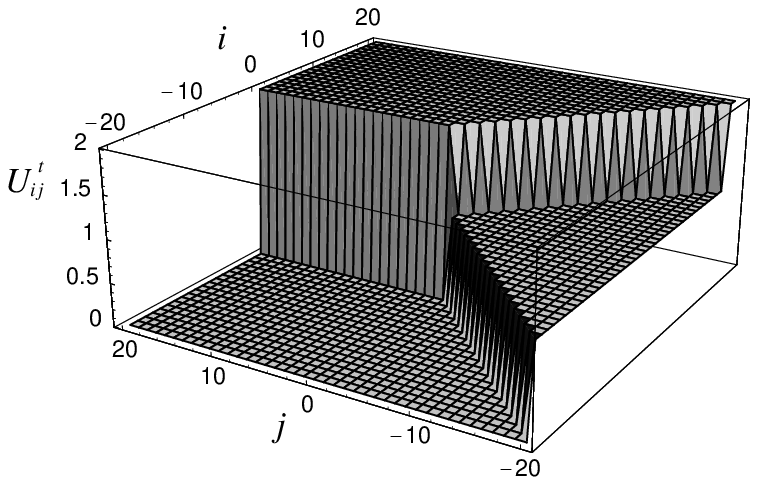}
\newpage
 \begin{center}
   {\bf Figure 2 by K.~Nishinari, et al}\\
 \end{center}
   \vspace{6cm}
   \epsfxsize=12cm    \epsfbox{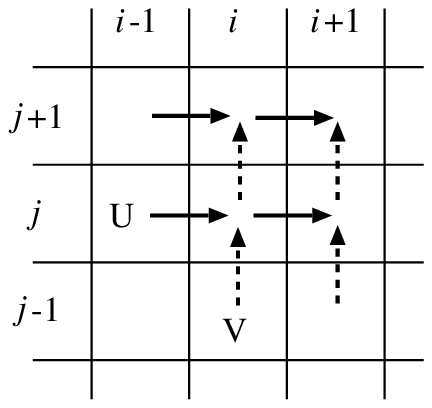}
\newpage
 \begin{center}
   {\bf Figure 3 by K.~Nishinari, et al}\\
 \end{center}
   \vspace{4cm}
   \epsfxsize=15cm    \epsfbox{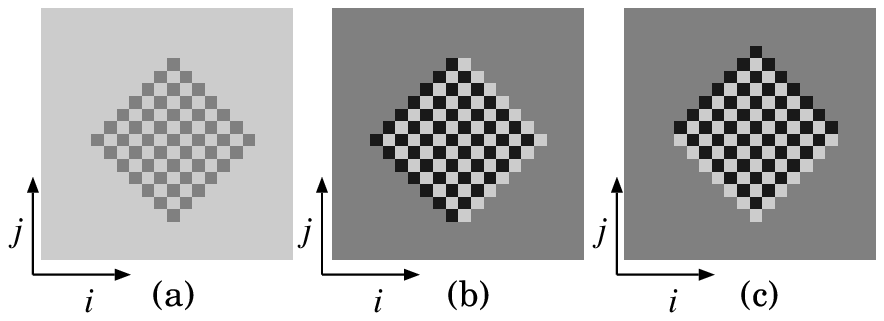}
\newpage
 \begin{center}
   {\bf Figure 4 by K.~Nishinari, et al}\\
 \end{center}
   \vspace{4cm}
   \epsfxsize=15cm    \epsfbox{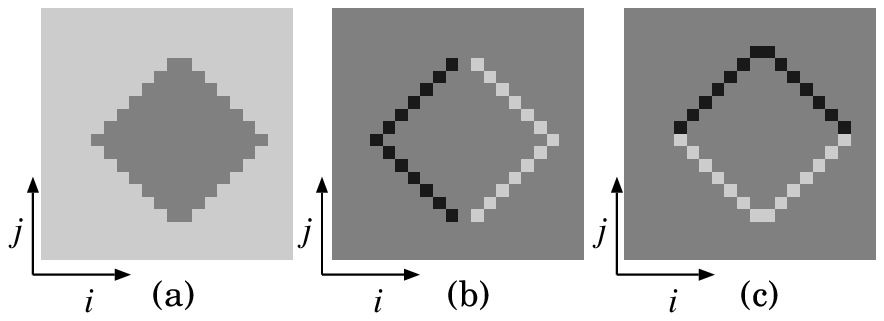}
\newpage
 \begin{center}
   {\bf Figure A1 by K.~Nishinari, et al}\\
 \end{center}
   \vspace{1cm}
   \epsfxsize=8cm    \epsfbox{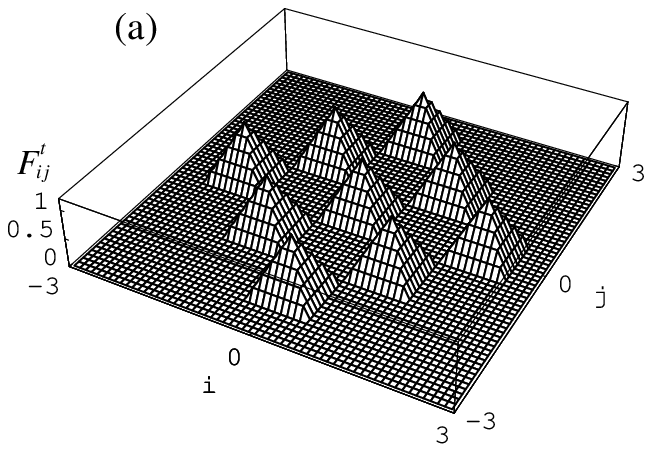} 
   \epsfxsize=8cm    \epsfbox{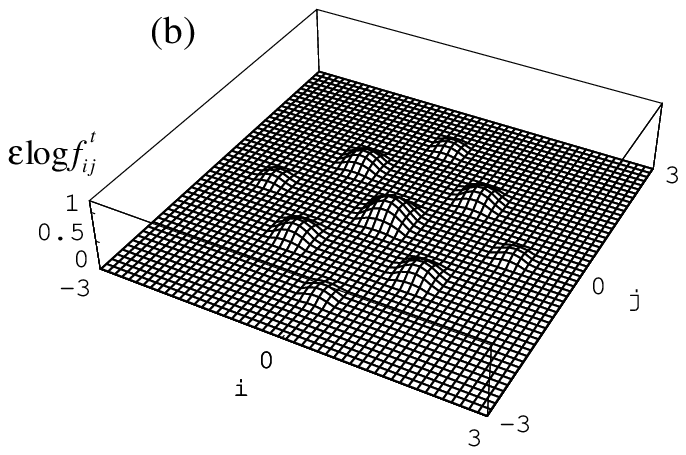} 
   \epsfxsize=8cm    \epsfbox{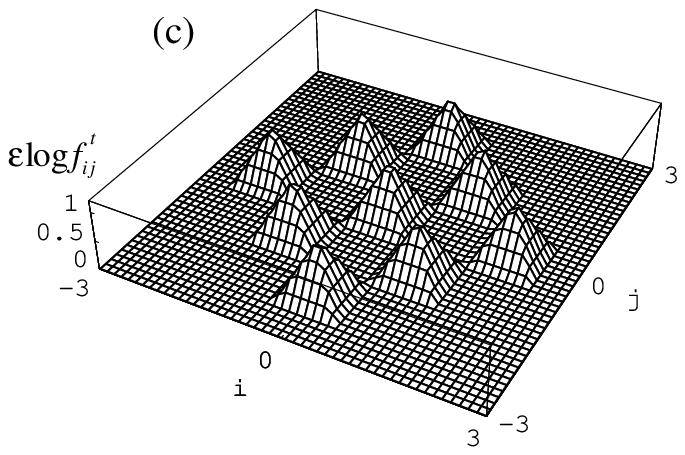}
}
\end{document}